\pdfoutput=1
\documentclass[conference]{IEEEtran}
\IEEEoverridecommandlockouts
\usepackage{cite}
\usepackage{amsmath,amssymb,amsfonts}
\usepackage{algorithmic}
\usepackage{graphicx}
\usepackage{textcomp}
\usepackage{xcolor}
\usepackage{siunitx}  
\usepackage{comment}

\usepackage{graphicx} 
\usepackage{subfigure}

\usepackage{multirow}

\usepackage{tikz}
\newlength\szg     \newcommand \hquan[1]{%
\settoheight\szg{#1}%
\tikz[baseline]{
\pgfmathparse{1}
\let\hfs\pgfmathresult
\filldraw (0,\szg/2) circle (\szg/2+0.35ex);
\node[white] at (0,\szg/2) {\makebox[0em][c]{\scalebox{\hfs}[1]{\textbf{#1}}}};
}}

\def\BibTeX{{\rm B\kern-.05em{\sc i\kern-.025em b}\kern-.08em
    T\kern-.1667em\lower.7ex\hbox{E}\kern-.125emX}}
\begin{document}

\title{SME: ReRAM-based \underline{S}parse-\underline{M}ultiplication-\underline{E}ngine to Squeeze-Out Bit Sparsity of Neural Network
}
\author{Fangxin Liu,
\and
Wenbo Zhao,
\and
Yilong Zhao,
\and
Zongwu Wang,
\and
Tao Yang,
\and 
Zhezhi He,
\and 
Jingnai Feng,
\and
Xiaoyao Liang,
\and 
Li Jiang*
\\
School of Electronic Information and Electrical Engineering, Shanghai Jiaotong University, Shanghai, China\\
\{liufangxin, zhaowenbo, zhezhi.he, sjtuj\}@sjtu.edu.cn, \{liang-xy, jiangli\}@cs.sjtu.edu.cn
}


\maketitle

\begin{abstract}
Resistive Random-Access-Memory~(ReRAM) crossbar is a promising technique for deep neural network~(DNN) accelerators, thanks to its in-memory and in-situ analog computing abilities for Vector-Matrix Multiplication-and-Accumulations~(VMMs). However, it is challenging for crossbar architecture to exploit the sparsity in the DNN. It inevitably causes complex and costly control to exploit fine-grained sparsity due to the limitation of tightly-coupled crossbar structure.

As the countermeasure, we developed a novel ReRAM-based DNN accelerator, named Sparse-Multiplication-Engine~(SME), based on a hardware and software co-design framework. First, we orchestrate the bit-sparse pattern to increase the density of bit-sparsity based on existing quantization methods. Second, we propose a novel weigh mapping mechanism to slice the bits of a weight across the crossbars and splice the activation results in peripheral circuits. This mechanism can decouple the tightly-coupled crossbar structure and cumulate the sparsity in the crossbar. Finally, a superior squeeze-out scheme empties the crossbars mapped with highly-sparse non-zeros from the previous two steps. We design the SME architecture and discuss its use for other quantization methods and different ReRAM cell technologies. Compared with prior state-of-the-art designs, the SME shrinks the use of crossbars up to $8.7\times$ and $2.1\times$ using Resent-50 and MobileNet-v2, respectively, with $\leq 0.3\%$ accuracy drop on ImageNet. 

\end{abstract}


\section{Introduction}

Resistive Random-Access-Memory~(ReRAM) crossbar emerges as a promising solution to accelerate the inference of Deep Neural-networks~(DNNs)~\cite{shafiee2016isaac,chi2016prime,song2017pipelayer}. ReRAM crossbar structure can efficiently carry out Vector-Matrix Multiplication~(VMM) -- the dominating computation of DNNs' inference~\cite{deng2020model} -- in the analog domain based on Kirchhoff's laws~\cite{shafiee2016isaac,chi2016prime}. The massive weights of DNNs are stationed at the ReRAM crossbar and reused as one of the operands in the VMM computation, significantly reducing massive data movements between on/off-chip memory. However, such a highly-parallel computing paradigm tightly couples the layout of the weight operands on ReRAM cells with the VMM calculation of the crossbar structure, making it difficult to exploit DNNs' sparsity. Consequently, the requirement of ReRAM crossbars to accommodate enormous DNN weights become the main hurdle of this technology. 

Extensive works propose weight sparsification methods to reduce the number of DNN weights. Hardware-independent pruning algorithms, such as filter-/channel-wise sparsity, can derive a ``smaller" dense weight matrices and directly map them to crossbars~\cite{liang2018crossbar}. These algorithms, however, are too coarse-grained and lead to limited sparsity utilization. Finer-grained sparsification methods adopt the software-hardware co-optimization fashion~\cite{wang2018snrram,ji2018recom,chu2019pimprune}. These works first map the weights to crossbars and then prune the whole crossbar-columns or crossbar-rows. The weight matrix change asks for extra peripheral circuits to coordinate the shape of input and output feature maps. Moreover, DNNs have to be retrained, which is not always feasible in a real-world scenario~\cite{deng2020model}. 


The second source of the vast requirement of ReRAM crossbars is derived from the small bit-width, i.e., 1-3 bits~\cite{Cai2019lowbit}, that a ReRAM cell can stably store due to process imperfections and limitation. Thus, quantization is necessary to reduce the bit-width of the weight. Most existing ReRAM-crossbar accelerators quantize the weight in 8-bits and decompose the 8 bits to 8 cells, respectively. The resulting massive \emph{bit-level sparsity} are difficult to exploit. A sparse ReRAM-crossbar architecture  SRE~\cite{yang2019sparse} attempts to exploit fine-grained sparsity derived by exchanging the crossbar-columns and crossbar-rows. This design can also exploit the bit-level sparsity. However, the hardware overhead due to the complex control, costly indexing, and routing almost offset the area-efficiency derived from the reduced crossbars.

The fundamental limit of exploiting the sparsity is because \textit{the data mapping and the VMM computation are tightly coupled with the crossbar structure}, denoted as \emph{structural-coupling problem}. To solve this problem, in this paper, we devote to exploit the bit-level sparsity to improve the area- and energy-efficiency of ReRAM-crossbar based DNN accelerators. We propose an algorithm-hardware co-design framework called SME by novel weight mapping schemes and data path design to squeeze out the bit-wise sparsity. SME can apply to many quantization methods, and it is training-free and orthogonal to existing pruning methods. The contributions are summarized as follows:
\begin{itemize}
\item We propose a bit-wise sparse pattern and an inter-crossbar bit-slicing scheme to accumulate the 0-bits to the same crossbars. 

\item We propose a squeeze-out scheme that empties highly sparse crossbars by sacrificing limited amount of least-significant bits.

\item We design the hardware architecture of SRE with a limited $2\,$Kb overhead. The proposed SME reduces up to $8.7\times$ and $2.1\times$ crossbars for Resent-50 and MobileNet-v2, respectively, compared with the SOTA method.

\end{itemize}

The reminder of this paper is as follows: Section II introduces the background and motivation; Section III - V demonstrate the SME algorithm, architecture, and experiments, in respect. Section VI concludes this paper.
\section{Background AND Motivation}
\subsection{ReRAM-based Sparse NN Accelerators}
\label{sec:sparse_nn_acce}

\begin{figure}[htb!]
\vspace{-0.5cm}
\setlength{\abovecaptionskip}{-0.10cm} 
    \centering
    \includegraphics[width=0.9\linewidth]{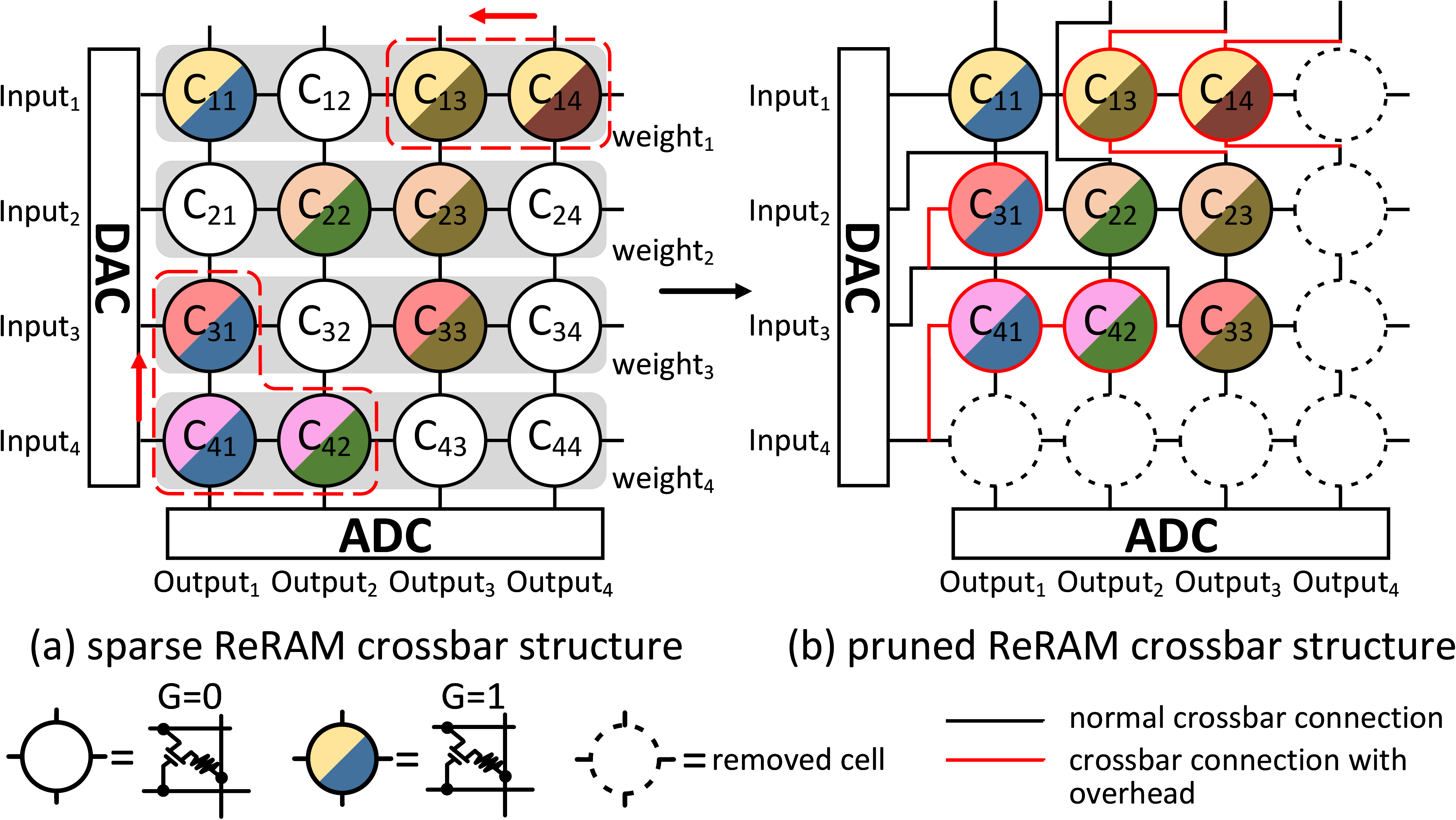}
    \caption{Crossbar row/column exchange and structural-coupling problem. The same color of the upper/lower semicircle indicates that the cells share the same input, and the cells' results are accumulated in the same bit-line, respectively.}
    \label{fig:reram}
    \vspace{-0.5cm}
\end{figure}

The \emph{Structural-coupling problem} manifests itself as the inability to freely skip the multiplication of zero operands because weight-bits in the same crossbar-row share the same input, and the current derived by multiplication in cells are accumulated in the same crossbar-column. In Fig.~\ref{fig:reram}(a), suppose each weight has 4-bit and is partitioned into four cells. If a single cell containing 0-bit is removed, other cells can not fill their position since they are from a different row or column. Moving weight-bits across crossbar-rows/columns leads to wrong MAC results as shown in Fig.~\ref{fig:reram}(b). 

Structural pruning methods avoid this problem by pruning the weights in a granularity that the whole crossbar-column~(or -row) can be removed at the cost of extra peripheral circuits~\cite{wang2018snrram,chu2019pimprune}. The extra peripheral circuits, including input-fetching and output-alignment modules, coordinate the feature maps to the pruned weight mapping on the crossbar. We break down the peripherals' area overhead and find PIM-Prune~\cite{chu2019pimprune} needs $4$KB index storage to skip fetching the unnecessary activation~(multiply by zero weight) for ResNet-50. It inevitably degrades the DNN's accuracy to prune a group of weights in such a granularity~(e.g., a $2\times 2$ kernel composed of ${weight}_{1-4}$ in Fig.~\ref{fig:reram}(a)). Thus, these works have to retrain the DNN to minimize the accuracy drop. 

The weight matrix after structural pruning still contain many 0-bit~(see $C_{21}, C_{24}$ in Fig.~\ref{fig:reram}(a)). As a result, enormous ReRAM-cells are mapped with 0-bits, denoted as sparse cells. SRE~\cite{yang2019sparse} breaks up crossbars into smaller parts, namely Operating Units~(OUs), to operate VMM independently. SRE exchanges the weight-bits' mapping on the OU-rows to cluster the zeros in OUs and empties the zero OUs for resource reduction. Thus, only the small OUs need to comply with the structural-coupling problem.  However, SRE fails to exploit sparsity in the column direction due to the dislocation problem~\cite{chu2019pimprune}. Moreover, SRE dramatically increases the peripheral circuit's overhead to accumulate the correct result and introduces $778$KB index storage for ResNet-50. There is a dilemma between the sparse utilization and the OU size: shrinking the size of OUs can exploit finer-grained sparsity but significantly increase the index overhead and the routing overhead of control circuits~(see the extra routing in Fig.~\ref{fig:reram}(b)).

\subsection{Weight Quantization for ReRAM-crossbar}
\label{sec:quantization}

The ReRAM cell is programmed into multiple conductance levels to represent a value, e.g., a 4-bit value requires $2^4$ different levels. The ReRAM cell's process limitations ~\cite{yang2019sparse} constrain the bit-width of the value a ReRAM cell can store. Consequently, a weight is conventionally segmented into multiple subwords; each subword is deployed on a ReRAM cell. This segmentation results in high crossbar costs. For example, ResNet-18 with 32-bit weights consumes more than $20,000$ crossbars of $128 \times 128$ size~\cite{Cai2019lowbit}. Thus, quantization is compulsory to reduce the bit-width of DNNs's weight. Quantization using weight clustering~\cite{wang2019haq} and sharing~\cite{li2019additive} strive to reduce the number of values that represent weights. The introduced index mechanism becomes an obstacle for ReRAM-crossbar. INT8 quantization~\cite{deng2020model} is thus an well-accepted quantization method for ReRAM-crossbar~\cite{yao2020fully}. The resulting integer values can be well aligned and mapped to ReRAM-crossbar. 

Recently, aggressive quantization methods are proposed to reduce the bit-width further. For example, POWER-OF-2 based quantization (PO2)~\cite{deng2020model,li2019additive} quantize values in the form of exponents. However, they incur great DNN finetuning effort and considerable accuracy drop~\cite{deng2020model}. Consequently, APT~\cite{li2019additive} quantizes weights into the sum of several power-of-twos with negligible accuracy loss. Adaptive quantization methods, like~\cite{wang2019haq}, quantize weights using different bit-widths to optimize the memory occupation and access. The quantized codewords resulted from these aggressive quantizations, however, need be aligned to the same bit-width -- by padding `0's to both MSB and LSB -- before mapping them on the crossbar (See ${weight}_{2}$ in Fig.~\ref{fig:reram}(a)). 
\subsection{Motivation for bit-wise sparsity exploitation}

\begin{figure}[b!]
    \centering
    \setlength{\abovecaptionskip}{-0.15cm} 
    \vspace{-0.6cm}
    \includegraphics[width=0.75\linewidth]{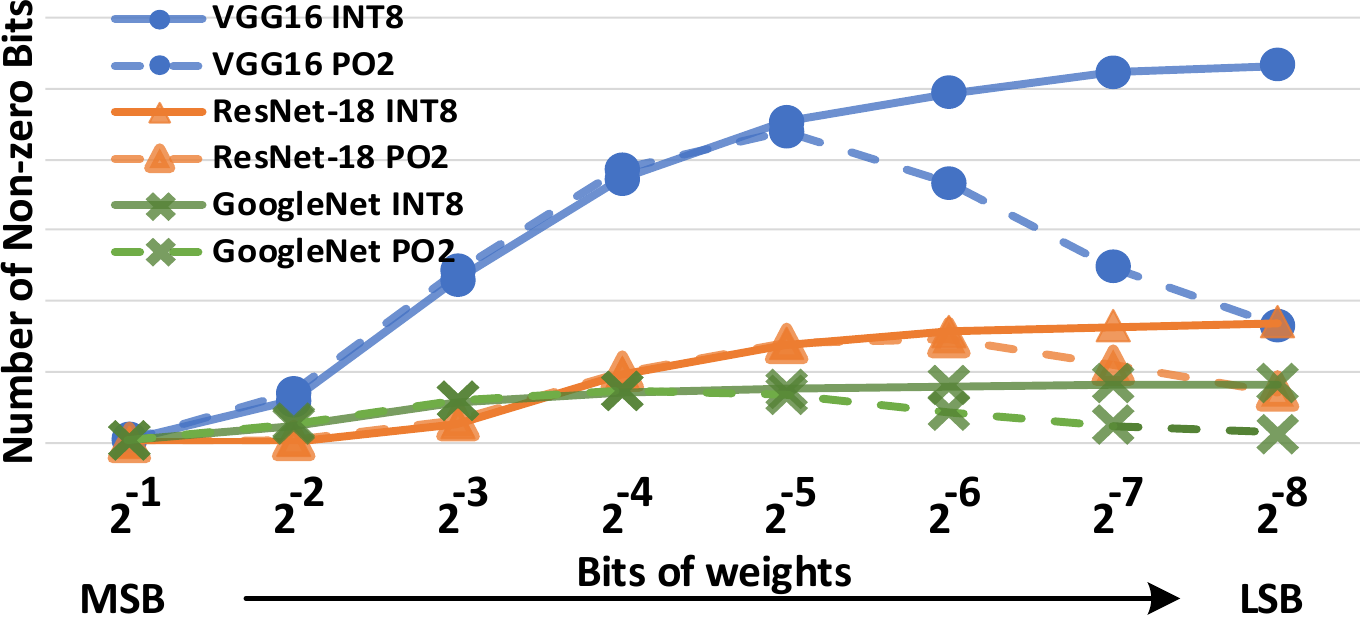}
    \caption{Comparison of the INT8 and PO2 quantization in terms of the bit-level sparsity of different weight bits in the quantized DNNs on ImageNet~\cite{deng2009imagenet}.}
    \label{fig:distribution}
    \vspace{-0.3cm}
\end{figure}

We observe the high bit-level sparsity conveyed in quantized weights. Fig.~\ref{fig:distribution} shows a significant sparsity in the first three most significant bits~(MSBs). In PO2 quantization, the last three least-significant bits~(LSBs) also contains many 0-bits. However, such bit-wise sparsity can hardly be exploited by ReRAM-crossbar accelerators due to the structural-coupling problem. The key is to decouple the crossbar structure, which is described in the next section.

\section{SME Algorithm}
\begin{figure}[btpb!]
\vspace{-0.1cm}
    \setlength{\abovecaptionskip}{-3pt}
    \centering
    \includegraphics[scale=0.12]{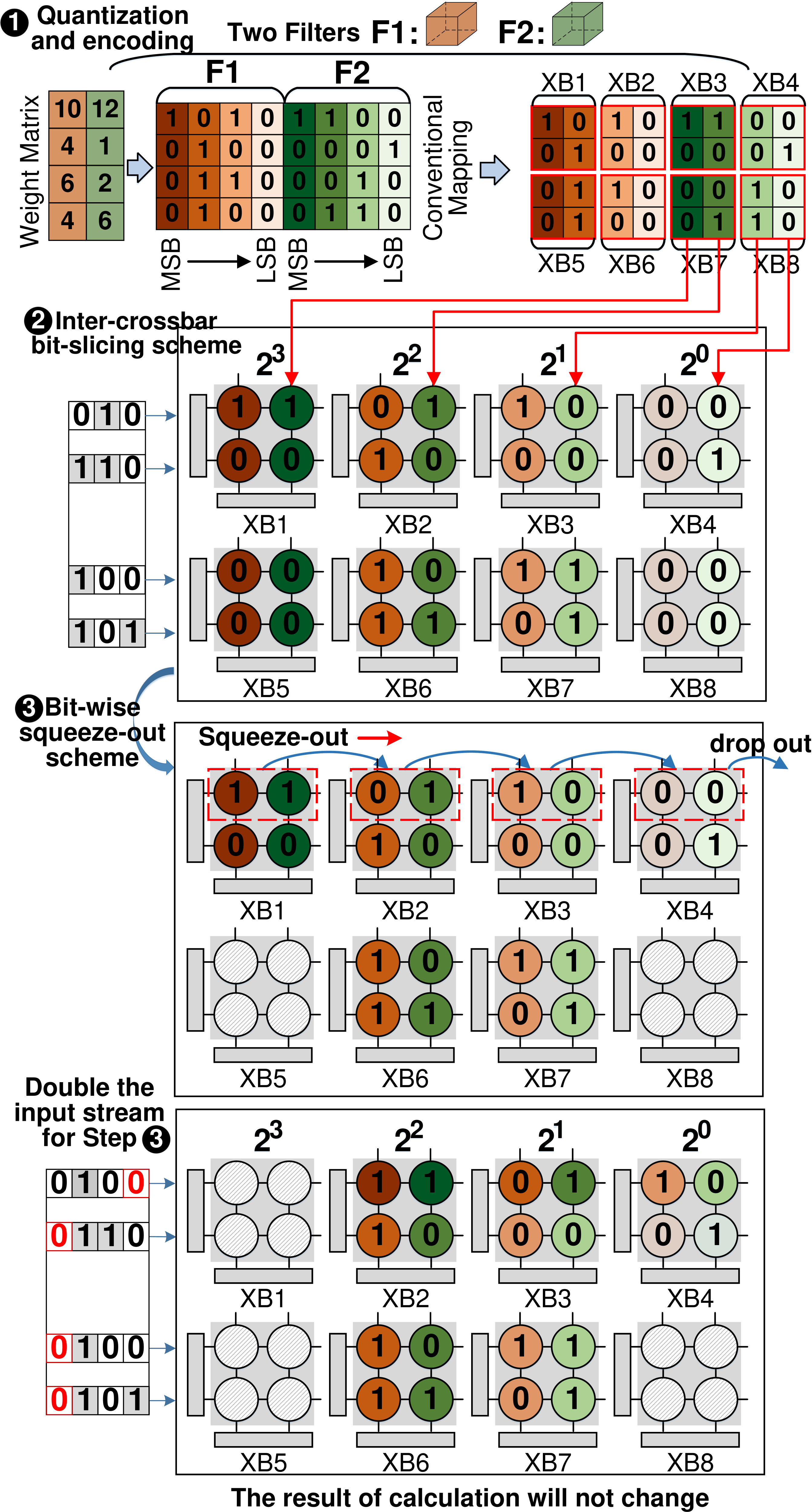}
    \caption{An example showing the process of SME.}
\label{fig:algorithm-framework}
\vspace{-0.5cm}
\end{figure}

We propose our SME scheme to decouple the crossbar structure by novel weight mapping algorithms. The SME mapping algorithm has three steps: quantization, inter-crossbar bit slicing and squeezing, as shown in Fig.~\ref{fig:algorithm-framework}.

\subsection{Quantization and Encoding Scheme}
\label{sec:mapping}
We assume single-level ReRAM cell as an example throughout this paper for simplicity. We quantize and encode the weights by extending the {APT quantization\footnote{Other quantization methods, such as adaptive quantization~\cite{wang2019haq,deng2020model} can also apply to SME, which is discussed in Section~\ref{sec:disscuss}.}}~\cite{li2019additive} that represents the weight with the sum of several power-of-twos. We map each $N_q$-bit weight onto $N_q$ cells, $b_{1:N_q}$. If the cell $b_i = 1$, the weight value contains a power-of-two $2^{-i}$. Then, the encoded $b_{1:N_q}$ can represent a quantized value:
\vspace{-0.3cm}
\begin{equation}
    w^q = \sum\limits_{i=1}^{N_q} b_i 2^{-i}, \quad b_i\in\{0,1\}
\vspace{-0.2cm}
\end{equation}    

\begin{figure}[tbp]
\setlength{\abovecaptionskip}{-0.2cm}   
\centering
\includegraphics[scale=0.2]{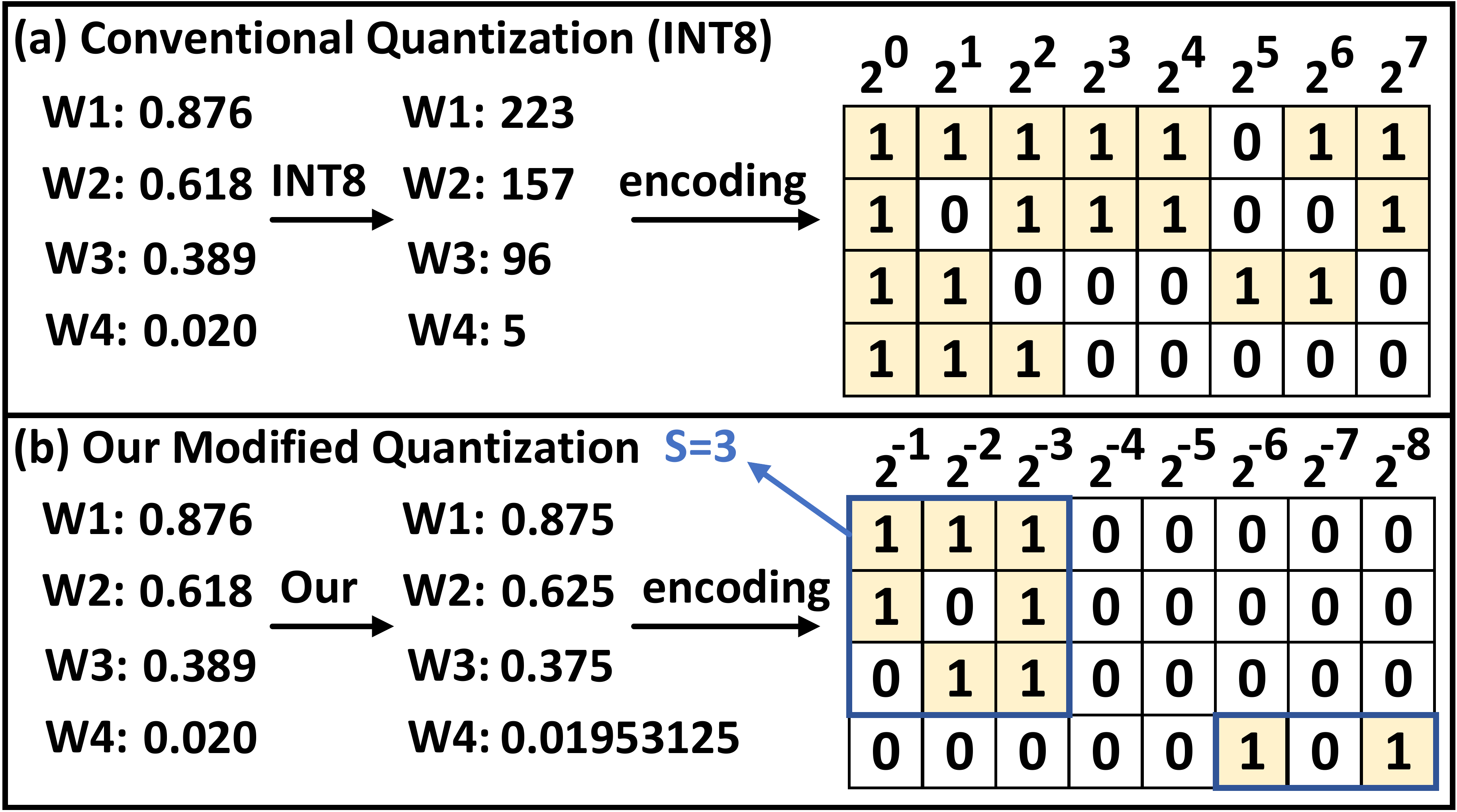}
\caption{The sparsity generated by INT8 and modified APT quantization.}
\label{fig:comparsion-int8}
\vspace{-0.5cm}
\end{figure}

In Step~\hquan{1}, we \emph{restrain the `1'-bits into a consecutive region of sizes $S$ in the binary codeword.} The above quantization can be derived by rounding the APT quantization result as follow:
\vspace{-0.2cm}
\begin{equation}
\begin{aligned}
    w^q = \sum\limits_{i=k}^{\min\{N_q,k+S-1\}} b_i2^{-i},\, k\in\{1,2,\ldots,N_q\}
\end{aligned}
\label{con:quan}
\vspace{-0.2cm}
\end{equation} 
The maximum absolute value Eq.~\ref{con:quan} can represent is $|w^q| \leq 1\!-\!2^{-S}$. Consequently, we scale all the weight value down into $\{\!2^{-S}-1\!, 1\!-\!2^{-S}\}$ using a simple shift operation in the architecture~(described in Section IV). Compared with the INT8 quantization, this quantization method can increase and accumulate the bit-level sparsity in a codeword, as shown in Fig~\ref{fig:comparsion-int8}.
\subsection{Inter-crossbar Bit-slicing Scheme}

To decouple the crossbar structure, we propose the \textbf{inter-crossbar bit-slicing} scheme. The key idea of \emph{bit-slicing} is to map the same bit of a quantized weight into the same bit crossbar, as shown in Step \hquan{2}. In this mapping process, a $W\!\times\! H$ weight matrix quantized with $N_q$ bits is sliced into $N_q$ bit-sliced matrices of size $W\!\times\! H$. Then, each bit-sliced matrix is further partitioned and mapped to ReRAM crossbars with size $xw \times xh$. $XB_i^1$ to $XB_i^{N_q}$, $i \in \{1, 2, \cdots, \lceil \frac{W}{xw} \rceil \times \lceil \frac{H}{xh} \rceil\}$ can also be viewed as the crossbar group, where the crossbar group is defined as the same region belonging to different bit-sliced matrices with the same size as the crossbar (such as $XB_{1-4}$ and $XB_{5-8}$ in Fig~\ref{fig:algorithm-framework}).

For example, in Fig~\ref{fig:algorithm-framework}, the first weight $1010$ in $F1$ is sliced and mapped onto the top-left cells of $XB_{1-4}$. MSBs in filter $F1$ and $F2$ is mapped to crossbar $XB1$ and $XB5$. The above mapping scheme can aggregate the sparsity in a crossbar~(e.g., $XB5$ and $XB8$). These empty crossbars can be saved by the mechanism of light-weight index~\cite{chu2019pimprune,zhao2020smartexchange,yang2019sparse}.



This bit-slicing mapping scheme requires minor modification on the peripherals~(refer to Section~\ref{sec:modules}), but it requires the same amount of peripheral circuits, such as ADCs, shifters, and adders, as conventional mapping method.
\begin{figure}[b!]
\vspace{-0.5cm}
\setlength{\abovecaptionskip}{-0.2cm}   
    \centering
    \includegraphics[scale=0.6]{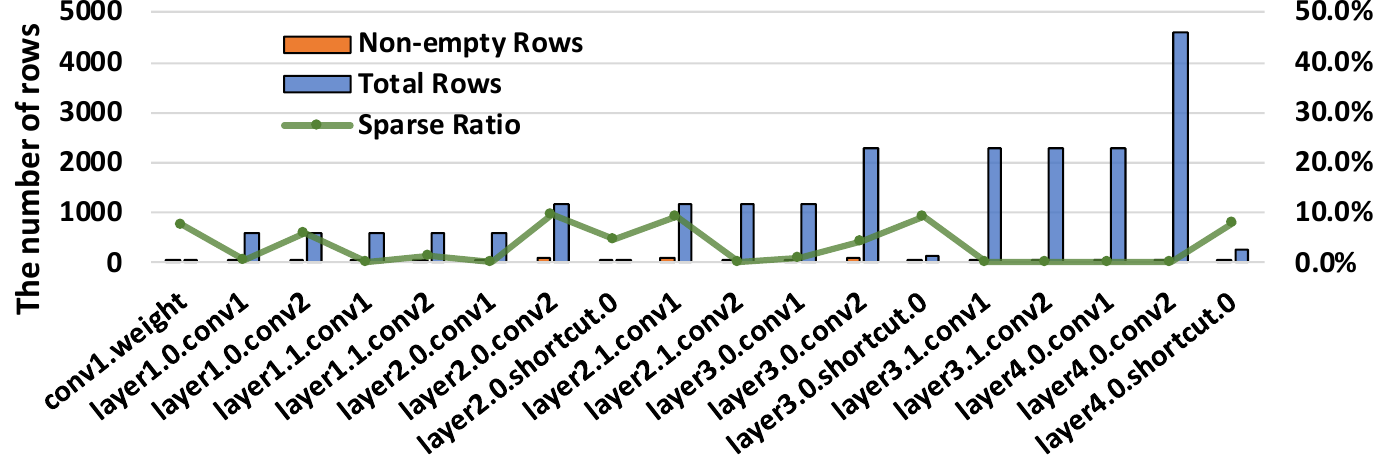}
    \caption{Non-empty row statistics of crossbars stored MSB of the ResNet-18.}
\label{fig:row-statistic}
\vspace{-0.5cm}
\end{figure}

\subsection{Bit-wise Squeeze-out Scheme}
\label{para: squeeze-out}
In the previous section, we aggregated a large amount of sparsity by the bit-slicing scheme so that some of the crossbars become empty and can be saved directly, but there are still crossbars that are very sparse that cannot be saved directly. We shrink the sparsity from the full crossbar to a smaller granularity, rows, since all the cells in the same crossbar-row share the same input. Fortunately, our SME approach make sure that the first few most-significant bit matrices are highly sparse. Fig.~\ref{fig:row-statistic} shows that less than 10\% non-empty rows in the most significant bit matrix in average. Based on above observation, we propose a clever squeeze-out scheme that circumvents the structure couple problem. The essence is row swapping among the crossbar group, but without introducing either overhead or large accuracy loss.

In Step~\hquan{3}, we squeeze the crossbar-rows containing non-zeros in preceding XBs to the subsequent XBs until these rows in tailing XBs are dropped out. For example, the first crossbar-row in XB$_1$ is remapped to XB$_2$, whose first crossbar-row is shifted to XB$_3$. And the LSB crossbar XB$_4$ drops its first crossbar-row. Based on step~\hquan{1},  release these crossbars will lead to no loss on network's accuracy. A corresponding operation on the input of these rows is performed. This step does not introduce extra indices for accumulating the output.

The following observations inspire the bit-wise squeeze-out scheme: the first few bit matrices are too sparse to compose a single crossbar. According to our bit-wise sparse pattern, `1's will only appear in successive positions after its most significant bit, which means that for the weights whose first few bits are `1', their last few bits must be `0'. We can empty crossbars that store the MSB by squeezing-out these bits without changing the actual quantized weight.

After squeeze-out by one bit, the
corresponding non-empty contents in $XB_i^j$ are moved to $XB_i^{j+1}$ ($j \in \{1,2,\cdots, N_q\}$) and these contents in the last bit crossbar $XB_i^{N_q}$ are abandoned. According to Eq.~\eqref{con:quan}, the value in that row is approximately halved with 1-bit squeezed. To ensure the invariance of the calculation results, we propose a scheme to double the input~(refer to Step \hquan{3} in Fig.~\ref{fig:algorithm-framework}). Note that we can even perform this step iteratively to squeeze-out multiple bits. 

\begin{figure}[htbp]
\vspace{-0.5cm}
\centering
\setlength{\abovecaptionskip}{-0.18cm}   
\includegraphics[scale=0.2]{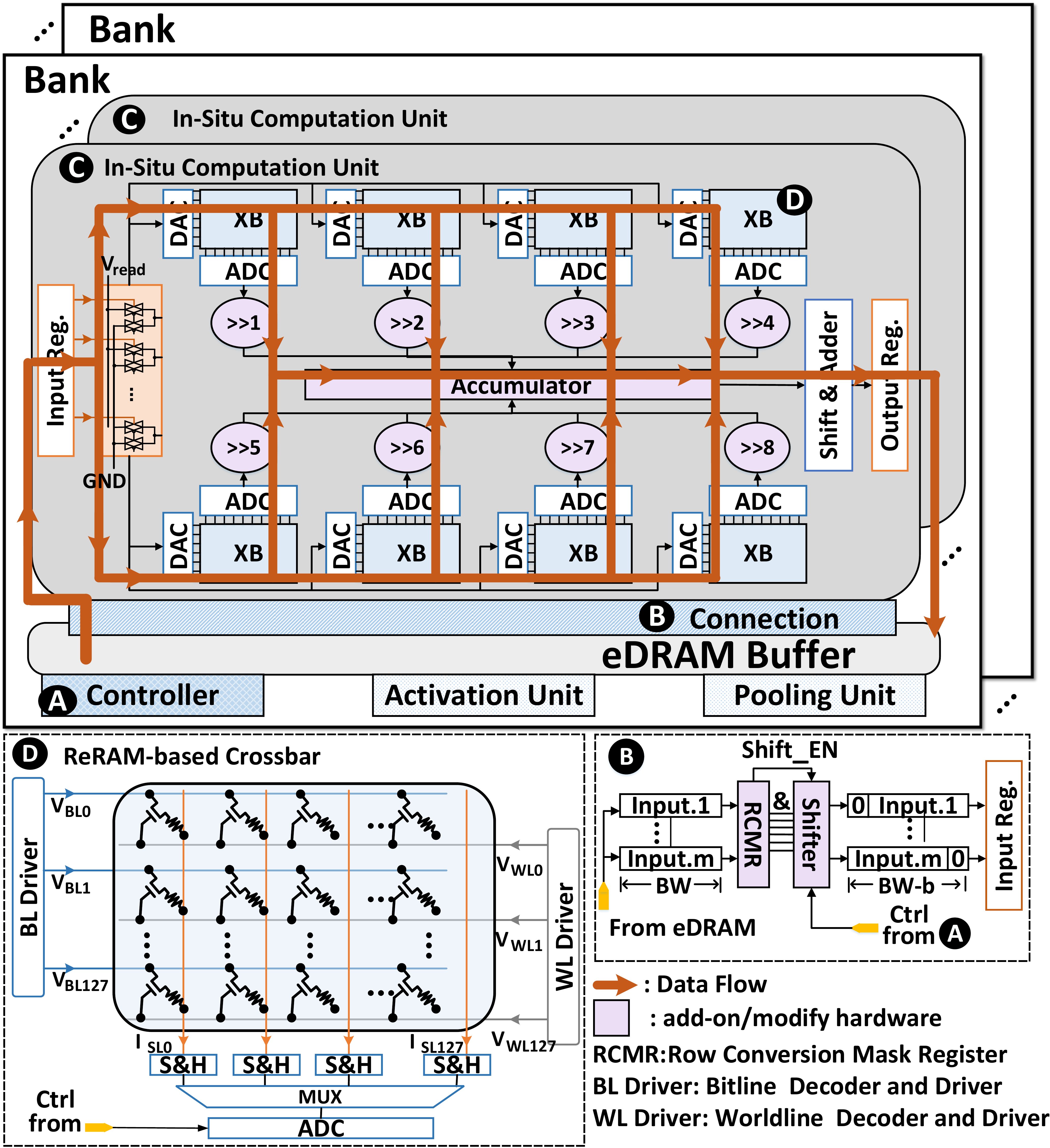}
\caption{Architecture Overview and Data Path of SME.}
\label{fig:architecture}
\vspace{-0.5cm}
\end{figure}

In this paper, we use the same style of input as in ISAAC~\cite{shafiee2016isaac}. The input is converted into bit-serial voltage and the number of cycles required is equal to the bit-width of the input.
If we squeeze out with $x$-bit, we delay the input of these rows for $x$ clocks, which implemented by Fig.~\ref{fig:architecture} \hquan{B}.


Assume that the weights and inputs are both 4-bit and we perform squeeze-out scheme for 1-bit.
For example, in Fig.~\ref{fig:algorithm-framework}~\hquan{2}, the first row in XB$_1$ is non-empty. Starting from XB$_1$, the crossbar's non-empty rows are placed at the same position in the latter crossbar, and the rows of the last crossbar (XB$_4$ in  Fig.~\ref{fig:algorithm-framework}) are dropped out. 
After that, the crossbar XB$_1$ storing MSB (i.e., $2^{-1}$) can be saved, and the bit-width of weights changes from $4$- to $3$-bit. The bits representing $2^{-1}$ in XB$_1$ is moved to XB$_2$ and represent $2^{-2}$, which means the part of weight shrinks by half (i.e., $W_1$ changes from $10 = 1010_{(2)}$ to $5 = 0101_{(2)}$). Thus, we shift the input of the first row one bit to the left ($input\times2$), which acts on the first row in each remaining crossbar (i.e., XB$_2\sim $XB$_4$) to make up for the changes in weight, i.e. $I_1\times W_1 =(I_1\times2)\times(W_1/2)$. This process equals to delaying the input one cycle, which means that the cycle of input changes from $4$ to $5$. As a result, the total amount of computation changes from $4\times H\times W\times 4$ to $5\times H\times W\times 3$, the computation has been reduced.

\section{SME Architecture}
\label{sec:architecture}
\subsection{Architecture Overview}
We present the overview of SME architecture, aiming at inference in edge devices. As shown in Fig.~\ref{fig:architecture}, each bank consists of three parts, all of which are connected to the shared bus:
1) the controller decodes instructions and provides control signals to all the peripheral circuits;
2) in-situ Computation Units~(CU) is the core computing and storage unit;
3) the shared blocks contain the activation unit, pooling unit, and eDRAM buffer for storing activations (i.e., intermediate computing results).
The SME add-on hardware implements the computation function of matching our algorithm, including simple modifications to the existing crossbar peripheral circuits, which is easier to manufacture than integrate complex logic into the chip.


\subsection{Module and DataFlow}
\label{sec:modules}
\textbf{Controller.}
Controller in Fig.~\ref{fig:architecture}~\hquan{A} provides control signals to all the peripheral circuits and drive the finite state machines that steer the inputs and outputs correctly after every cycle based on the technique configurations. 

\noindent\textbf{In-situ Computation Unit.}
Fig.~\ref{fig:architecture}~\hquan{C} shows the CU that is composed of crossbars and peripheral circuits. It includes DAC/ADCs for data conversion, accumulator and shifters, which sum up the partial sums of crossbars, and shift-adder unit for aggregating the partial results of input cycles.

First, the input is in the form of the bit sequence, and then each cycle enters respectively $8$ crossbars to the same in-situ CU to carry out MAC operation, and each gets the output currents.
After output current generated on SL, first, latch the computation results by S\&H circuits, and then sent $128$ analog voltages to ADC to convert them into digital signals through a MUX. In each cycle, $8$ crossbars are sampled in parallel, and a calculation result of different bits of the same weights is obtained through the corresponding shifter is connected to each crossbar. Then, these results are sent to the accumulator to generate the complete calculation results of the weight.
Finally, the results transmit into the output register for updating the outputs by shifting and adding. After traversing the 128 samplings, the next cycle calculation is repeated. 
\noindent\textbf{ReRAM-based crossbar.}
Fig.~\ref{fig:architecture} \hquan{D} shows the ReRAM-based crossbar with $128 \times 128$ size, which perform parallel MAC operations. We adopt the SLC as the ReRAM cell since SLC is more reliable against process variation compare to the MLC counterpart. The ReRAM array is implemented with a one-transistor–one-memristor (1T1R) cell structure.


\noindent\textbf{Buffer Connection}, which supports the squeeze-out strategy.
Fig.~\ref{fig:architecture}~\hquan{B} shows the communication between eDRAM Buffer and input register. The RCMR is used for fetching inputs from buffer to the register, which needs to convert the bit-width of input enable increase input cycle caused by the squeeze-out strategy (refer to Section~\ref{para: squeeze-out}). Data from the buffer pass through the RCMR and the controller determines whether the squeeze-out scheme needs to be performed at the current layer, and if it does, the inputs bit-width will be extended to $(8 + x)$-bit (we initially default the weights were quantized to 8-bit). The shifter follows the RCMR, `1' means to shift $x$-bit to the left, while `0' indicates padding $x$ zeros in front of MSB. After that, data transmit into the input register.
\section{EXPERIMENTS}
\subsection{Experimental Setup}
\begin{table}[h]
\vspace{-0.7cm}
\setlength{\abovecaptionskip}{-0.1cm}  
\centering
\caption{Configurations of GEM5 used in our simulation.}
\label{tab:gem5_config}
\setlength{\tabcolsep}{4mm}{
\begin{tabular}{|c|c|}
\hline
\begin{tabular}[c]{@{}c@{}}On-Chip Buffer\\ (DRAM)\end{tabular}  
& \begin{tabular}[c]{@{}c@{}}HBM$\_$1000$\_$4H$\_$1$\times$128 model;\\ 16KB/Bank
\end{tabular}        
\\ \hline
Bank IO Bus                                                       & 42.56MB/s                                                                                  \\ \hline
\begin{tabular}[c]{@{}c@{}}ReRAM-based\\ Main Memory\end{tabular} & \begin{tabular}[c]{@{}c@{}}20MB ReRAM; 100ns/cycle\\ 8 crossbars/CU;\   8 CUs/bank;\end{tabular} \\ \hline
\end{tabular}
}
\vspace{-0.3cm}
\end{table}
We adopt the configurations similar to the analysis ISAAC and modify GEM5~\cite{binkert2011gem5} to build a simulator for ReRAM-based crossbar architecture using the same configuration as ISAAC~\cite{shafiee2016isaac}. The configurations of DRAM, Bank, and ReRAM main memory are shown in Tab.~\ref{tab:gem5_config}. We redesign the data path since we add some peripheral circuits to integrate the techniques. Moreover, the energy consumption and area overhead of memories, including eDRAM buffer, input/output register, and registers stored mask (i.e., RCM Register), are calculated with CACTI~\cite{balasubramonian2017cacti} based on the 32-nm CMOS process. The memristors adapted SLC with a resistance range of $10K\Omega \sim 100K\Omega$ and crossbar size set as $128\times128$. 

We use the ISAAC as the baseline. We evaluate our work on classical image classification task, using several representative DNNs (ResNet-18/50~\cite{he2016resnet}, MobileNet-v2~\cite{sandler2018mobilenetv2}) on ImageNet ILSVRC-2012~\cite{deng2009imagenet}.
We compare with the PIM-Prune~\cite{chu2019pimprune}, SRE~\cite{yang2019sparse}, SmartExchange~\cite{zhao2020smartexchange}. For results that not available, we reproduce their experiments and report the results. Our method can also combine with other sparsity utilization solutions, such as SRE, PIM-Prune, etc. We implement our SME algorithm framework in the Pytorch framework to valid it. 
\subsection{Results and Analysis}
\emph{1) Accuracy and Sparsity: } Tab.~\ref{tab:acc} first shows the NN accuracy for the networks on ImageNet datasets. We can observe that our SME and other solutions are orthogonal and can further improve the effect combines with the exciting methods with negligible accuracy loss. Specifically, for ResNet-50, SME combines with PIM-Prune achieve $91.23\%$ sparse rate with $0.6$ accuracy loss compared to SmartExchange ($58.6\%$ sparse rate with $2.07\%$ accuracy loss) and PIM-Prune ($71.91\%$ sparse rate with $1.10\%$ accuracy loss) on ImageNet.
\begin{table}[htbp]
\vspace{-0.3cm}
\setlength{\abovecaptionskip}{-0.1cm}  
\centering
\caption{Inference accuracy on ImageNet}
\label{tab:acc}
\setlength{\tabcolsep}{1mm}{
\begin{tabular}{ccccc}
\hline
Model & Ori.Acc (\%) & Method & Acc.(\%) & Sparsity (\%) \\ \hline
\multicolumn{5}{c}{ImageNet ILSVRC-2012} \\ \hline
\multirow{4}{*}{ResNet-50} & 76.13 & SmartExchange & 74.06 & 58.60 \\
 & 76.13 & PIM-Prune & 74.91 & 71.91 \\
 & \textbf{76.13} & \textbf{SME} & \textbf{76.03} & \textbf{67.35} \\
 & \textbf{76.13} & \textbf{SME+PIM-Prune} & \textbf{75.46} & \textbf{91.23} \\ \hline
\multirow{4}{*}{MobileNet-v2} & 72.19 & SmartExchange & 70.16 & 79.79 \\
 & 71.88 & PIM-Prune & 70.11 & 77.13 \\
 & \textbf{71.88} & \textbf{SME} & \textbf{71.57} & \textbf{78.74} \\
 & \textbf{71.88} & \textbf{SME+PIM-Prune} & \textbf{71.02} & \textbf{84.51} \\ \hline
\end{tabular}
}
\vspace{-0.3cm}
\end{table}

\begin{figure}[ht]
\vspace{-0.2cm}
\setlength{\abovecaptionskip}{-0.2cm}   
\centering
\includegraphics[scale=0.46]{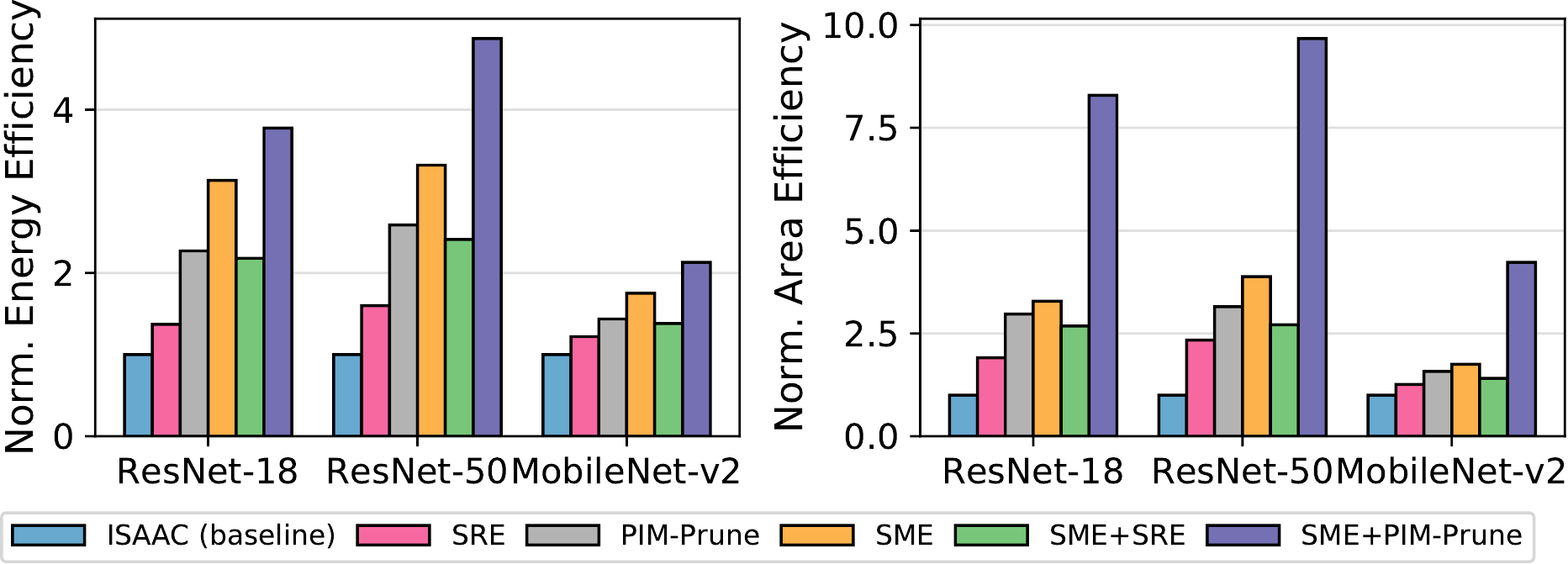}
\caption{Normalized energy- and area-efficiency.}
\label{fig:energy-efficiency}
\vspace{-0.3cm}
\end{figure}

\emph{2) Energy- and Area-Efficiency: } Fig.~\ref{fig:energy-efficiency} shows the energy- and area-efficiency of different accelerators for the four networks. We normalize the energy-efficiency to that of the model without any compression. On average, for ImageNet, SME improves energy efficiency by $2.3\times$ and area efficiency by $6.1\times$ on ResNet-18/50 compared to PIM-Prune and SRE. Even on MobileNet-v2, our method is still superior to the existing methods. The reason is that the pruning-based methods are difficult to compress networks for large-scale datasets or compact networks with acceptable accuracy. However, massive bit-level sparsities always exist and can be used by our SME.

\begin{figure}[ht]
\vspace{-0.3cm}
    \centering
    \setlength{\abovecaptionskip}{-0.2cm}   
    \includegraphics[scale=0.4]{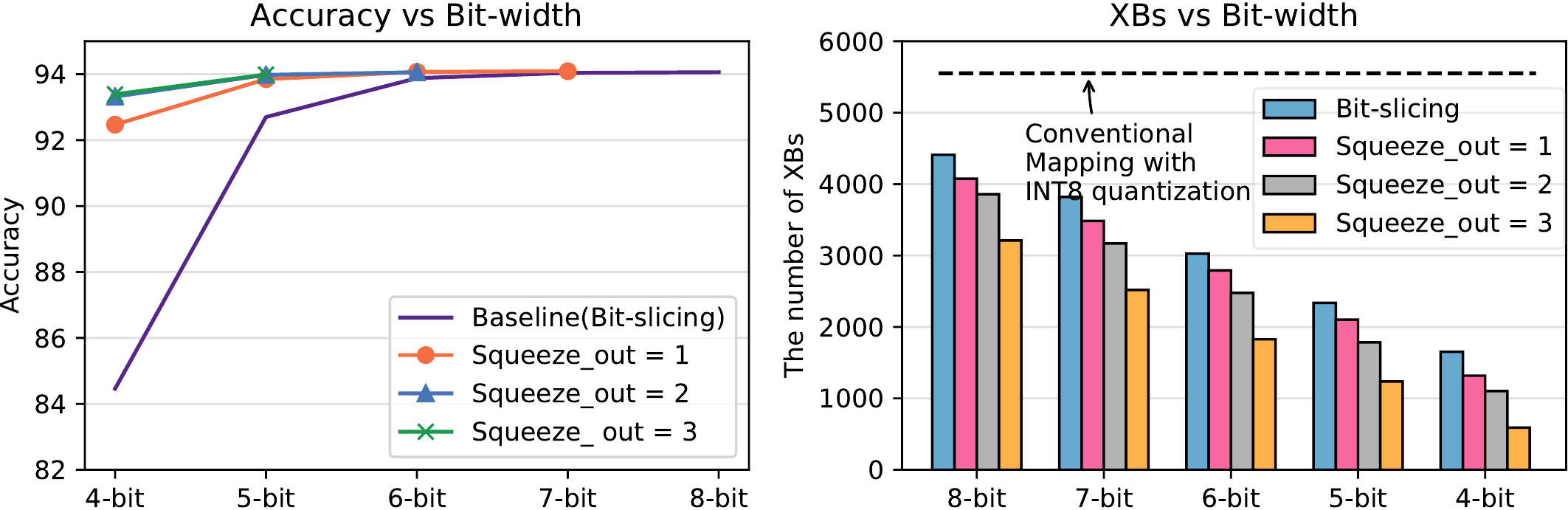}
    \caption{Comparison of NN accuracy and the number of crossbars with our different scheme on ResNet-18. Baseline is quantized with INT8 method.}
\label{fig:ACC_XBs}
\vspace{-0.2cm}
\end{figure}

\emph{3) Varied squeeze-out schemes with crossbar resource: } Fig.~\ref{fig:ACC_XBs} reports the results with the different squeeze-out schemes. We respectively compare the accuracy and necessary crossbar resources of squeezing 1,2,3 bits. We use the squeeze-out scheme to reduce the number of cells representing weights far better than directly reducing because the MSBs are more critical than the LSBs. If we can reduce the error caused by the MSBs, the overall error can be effectively decreased~\cite{Cai2019lowbit}.

\begin{figure}[ht!]
\vspace{-0.3cm}
\setlength{\abovecaptionskip}{-0.15cm}   
\centering
\includegraphics[scale=0.42]{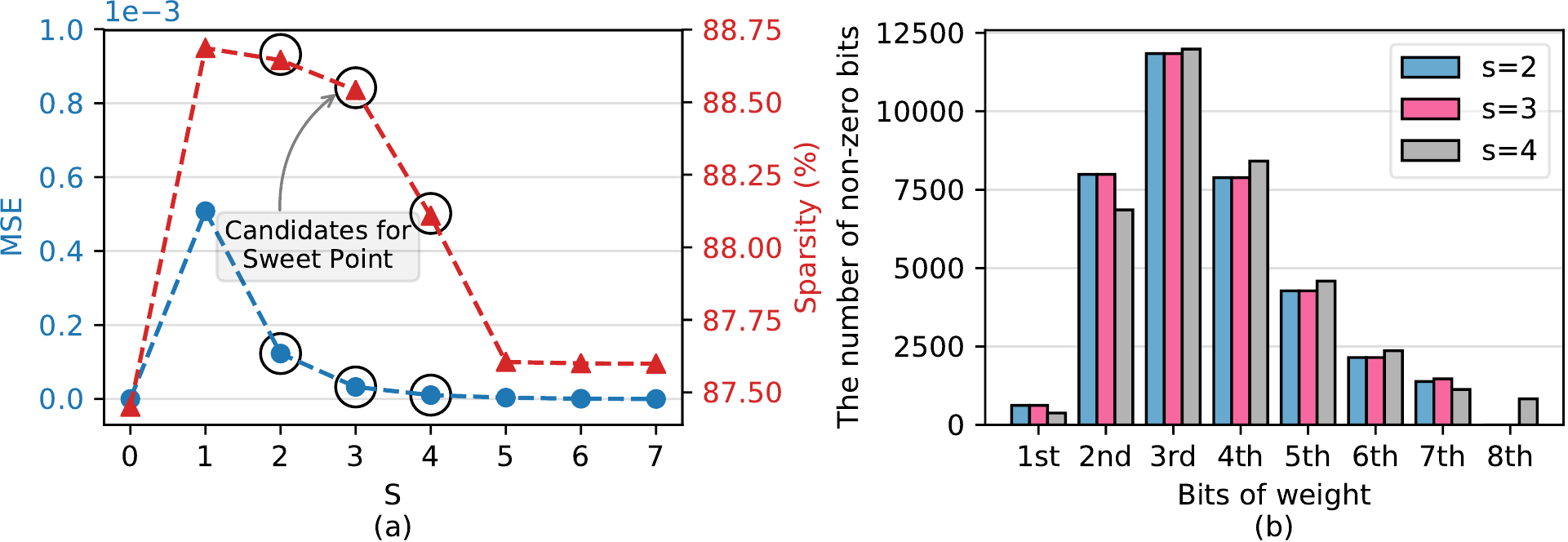}
\caption{(a) Trade-off between quantization error and bit-level sparsity along with the change of the consecutive region's size  containing `1.' w.r.t. $S$. (b) The sparsity of bit-level under different $S$ in the candidate set for sweet point.}
\label{fig:sparsity-mse}
\vspace{-0.3cm}
\end{figure}

\emph{4) Sweet-spot for the size of consecutive region containing `1': } As our discussion in section~\ref{sec:mapping}, Fig.~\ref{fig:sparsity-mse} shows the trade-off between the sparsity and quantization error caused by different number $S$ of consecutive `1'. We use mean square error (MSE) to measure the loss caused by quantization, which is defined as the absolute difference between the exact and the approximate weights~\cite{deng2020model}. 
In Fig.~\ref{fig:sparsity-mse}, we find if we set $S=2$, the overall sparsity of the network began to decrease, while $S=4$, the overall error of the model, is almost zero. However, we combine with the overall sparsity and the bit-level sparse distribution. We can find that $S=3$, SME achieves an optimal point for ResNet-18.

\begin{figure}[ht]
\vspace{-0.2cm}
\setlength{\abovecaptionskip}{-0.15cm}   
\centering
\includegraphics[scale=0.65]{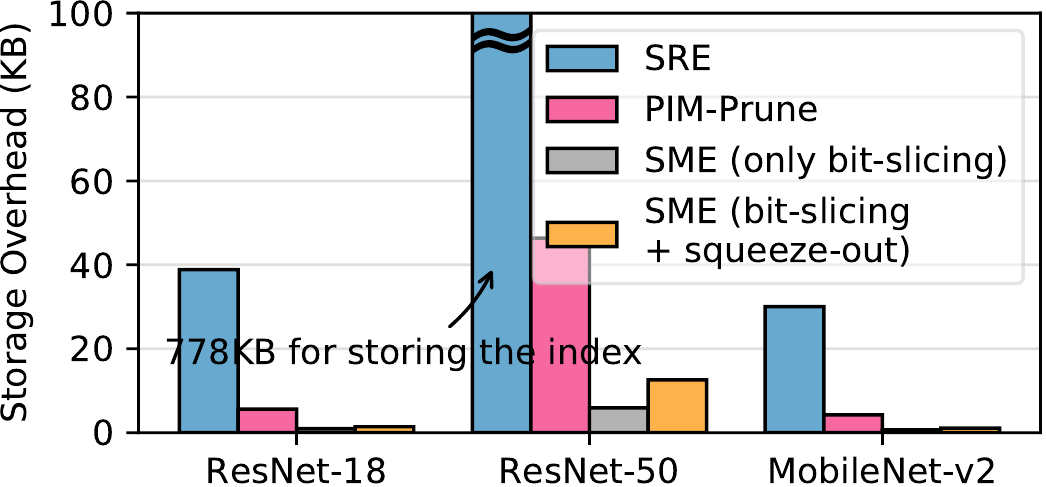}
\caption{Comparison of storage overhead with different networks.}
\label{fig:network-overhead}
\vspace{-0.3cm}
\end{figure}

\emph{5) Overhead Analysis: } Fig.~\ref{fig:network-overhead} shows the storage overhead of different networks. On average, SME achieves $84.6\%$ and $98.1\%$ register overhead reduction compared to PIM-Prune and SRE with only quantization and bit-slicing scheme; achieves $77.8\%$ and $96.8\%$ register overhead reduction compared to PIM-Prune and SRE further combining with the squeeze-out scheme. However, the significant reduction in overhead benefits from two parts: (1) our squeeze-out scheme solves the index's problem for aligning the output by processing the input. (2) we retain crossbars if the crossbar cannot be released, so the index is continuous.

\subsection{Design Exploration}
\label{sec:disscuss}
In this section, we discover our method can support the network with intra-layer mixed-precision~\cite{wang2019haq}, and also support MLC-based crossbar but perform better on SLC.
\begin{figure}[h!]
\vspace{-0.3cm}
\setlength{\abovecaptionskip}{-0.2cm}
\centering
\includegraphics[scale=0.6]{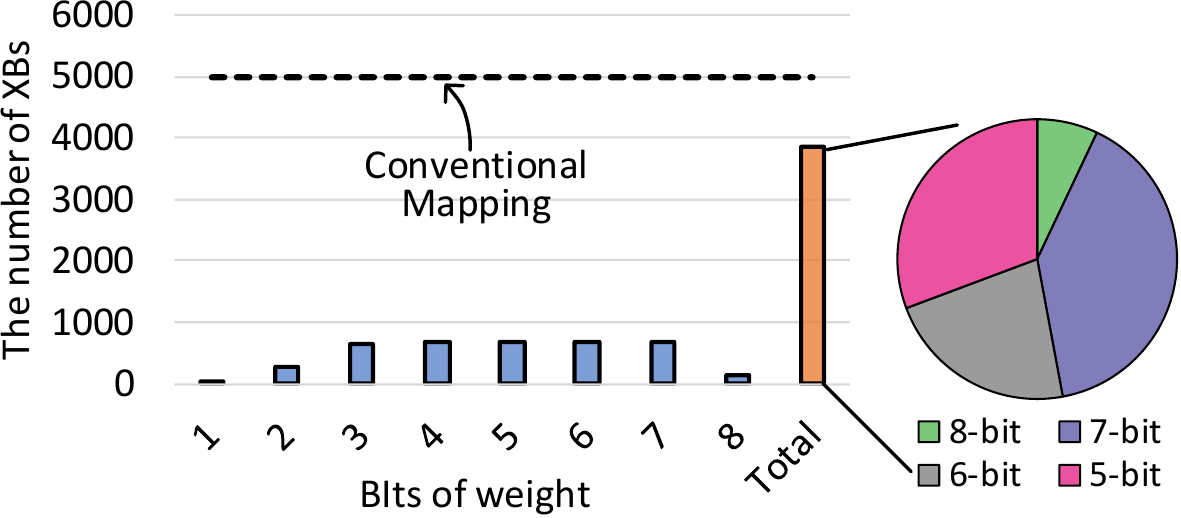}
\caption{\small The number of crossbar for ResNet-18 with mixed-precision. }
\label{fig:mixed-precision}
\vspace{-0.2cm}
\end{figure}


\emph{1) Support intra-layer mixed-precision:} Fig.~\ref{fig:mixed-precision} shows the crossbar ($128\times128$) consumption of ResNet-18 quantized by intra-layer mixed-precision under conventional mapping and our SME method. The pie chart of the Fig.~\ref{fig:mixed-precision} also shows the mixed-precision contains 5 to 8-bit in layers of ResNet-18. The conventional mapping approach cannot take advantage of mixed-precision benefits due to structural-coupling. The weights within a filter are mapped to crossbars, and the maximum bit-width of weights determines the number of cells required per weight. So there are massive sparse cells. In contrast, the SME slices each bit of the weights into different crossbars, aggregating the bits' sparsity. This achieves decoupling of the crossbar structure, reducing over $1,000$ crossbars than the conventional mapping method.
\begin{figure}[h!]
\vspace{-0.3cm}
\setlength{\abovecaptionskip}{-0.2cm}
\centering
\includegraphics[scale=0.9]{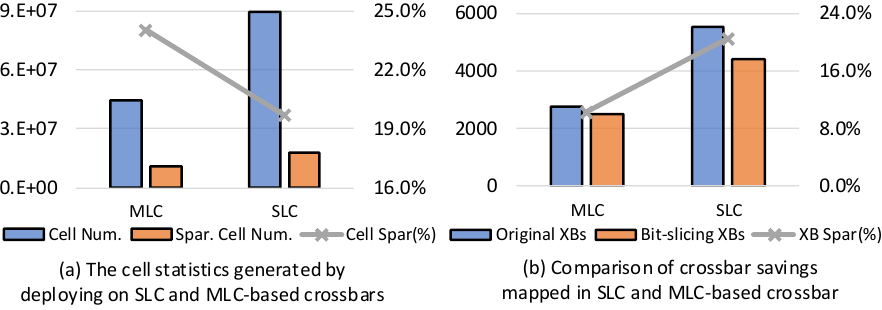}
\caption{\small Comparison of ResNet-18 deployed on SLC/MLC-based cell. }
\label{fig:MLC-map}
\vspace{-0.3cm}
\end{figure}

\emph{2) Support MLC-based crossbar:} Fig.~\ref{fig:MLC-map} shows that our bit-slicing scheme is also applicable to MLC-based crossbar. The number of sparse cells is significantly decreased when the network is mapped onto the MLC-based crossbar. Thus, our bit-slicing scheme's benefit is also affected, as we reduce approximately $11\%$ crossbars compared to the conventional mapping scheme. Besides, we can use the squeeze-out scheme to save resources further, and squeeze one MLC-based cell is equivalent to squeezing 2-bit on the SLC-based crossbar.


\section{Conclusion}
Bit-level sparsity cannot be utilized, leading to the limited performance of NN inference. We propose \textbf{SME}, an algorithm-hardware co-design framework that decouples the hardware dependence of multiplication to release the sparse cells in the crossbars for higher energy-/area-efficient inference of NNs. Besides, we design the architecture to efficiently support our algorithm through well-designed crossbars with the peripheral circuit.
Our evaluation shows that the proposed SME outperforms other similar solutions in energy, area, and accuracy.

\end{document}